\documentstyle[preprint,epsfig,aps,amssymb,amsmath,endfloat]{revtex}

\tightenlines

\begin{document}
\title{
\textsf{
\begin{center}
Molecular Weight Dependence of Polymersome \\
Membrane Structure, Elasticity, and 
Stability
\end{center}
}
}

\author{
\textsf{
\begin{center}
Harry Bermudez$^\ast$,
Aaron K. Brannan$^\dag$,
\\
Daniel A. Hammer$^\ast$,
Frank S. Bates$^\dag$,
and
Dennis E. Discher$^{\ast,}$
\footnote[3]{To whom correspondence
should be addressed: School of Engineering and Applied Science, 112 Towne
Bldg., University of Pennsylvania, Philadelphia PA 19104.
Phone: 215-898-4809, Fax: 215-573-6334, E-mail: discher@seas.upenn.edu}
\end{center}
}
}

\address{
\textsf{
\begin{center}
$^\dag$School of Engineering and Applied Science, University of
Pennsylvania, Philadelphia PA 19104,\\
$^\ddag$Department of Chemical Engineering and Materials Science, University
of Minnesota, Minneapolis MN 55455
\end{center}
}
}
\maketitle

\vskip 6.0cm

\textsf{
\begin{flushleft}
Classification: Physical Sciences - Applied Physical Sciences\\
Manuscript Information: 18 pages \\
Word and character counts: 195 words in abstract and 38,958 characters in paper \\
\end{flushleft}
}

\vfill\eject

\begin{abstract}
\textbf{\textsf{Vesicles prepared in water from a series of
diblock copolymers - ``polymersomes" - are physically characterized and
compared to lipid vesicles.
With increasing molecular weight $\bar{M}_n$, the hydrophobic core thickness
$d$ for the self-assembled bilayers of poly(ethylene oxide)-polybutadiene
(PEO-PBD) increases up to $\simeq$20 nm -
considerably greater than any previously studied lipid system.
The mechanical responses of these membranes, specifically, the area elastic
modulus $K_a$ and maximal areal strain $\alpha_c$ are measured by
micromanipulation.  As expected for interface-dominated elasticity, $K_a$
($\simeq$100 pN/nm) is found to be independent of $\bar{M}_n$, but lower
than the usual values for zwitterionic lipid membranes.
Experiments on polymersomes show $\alpha_c$ increases in a nearly linear
fashion with $\bar{M}_n$, approaching a limiting value predicted by 
mean-field ideas which
is universal and about 10-fold above that typical of lipids.  
Nonlinear responses and memory effects generally emerge with increasing
$\bar{M}_n$, indicating the onset of chain entanglements 
at higher $\bar{M}_n$.  
The effects of $\bar{M}_n$ thus suggest a compromise
between stability and fluidity for biomembranes. 
More generally, the results highlight the interfacial limits of 
self-assemblies at the nanoscale.
}
}\\
\end{abstract}

\vfill\eject

Biological systems of all sorts have long been appreciated as 
exploiting aqueous self-assembly; synthetic amphiphiles of many types 
have also been shown to spontaneously self-assemble 
into highly ordered structures in water 
\cite{nolte,meijer,forster,meier,kaler,discherfamily}.
Depending on temperature and molecular characteristics such as 
geometry, numerous morphologies are now possible,
including vesicles, micelles, and more exotic structures. 
Even so, the factors contributing to microphase stability 
are not always clear at the nanoscale, where interfacial effects often 
dominate bulk interactions.
Towards addressing these issues we 
describe the material properties of vesicle membranes made from a 
novel molecular weight series of diblock copolymers.

Lipid vesicles or ``liposomes" \cite{bangham} are often considered the 
prototypical membrane systems. 
As such, they have received considerable attention
that has proven relevant to the fundamentals of membrane behavior 
and to the motivation for biomimics \cite{lipowsky}. 
However, practical applications 
involving liposomes have been continually hindered by a lack of 
stability \cite{lasic}. 
Presumably commensurate with limits on membrane stability is the
narrow range ($3-5$ nm) of the hydrophobic core thickness $d$ 
of liposome membranes \cite{marsh}.
We are able to extend the range of $d$ and explore the impact on
membrane properties by forming vesicles from diblock copolymers of 
poly(ethylene oxide)-polybutadiene (PEO-PBD) \cite{wonmicelles}. \\

\begin{table}
\caption{Diblock copolymers and membrane core thickness $d$ examined here.}
\begin{tabular}{ccddd}
Designated&Polymer&$\bar{M}_n$&&$d$ \\
Name&Formula&(kg/mol)&$f_{EO}$&(nm)  \\
\tableline
\textbf{OE7}&EO$_{40}$-EE$_{37}$&3.9&0.39&8.0$\pm$1 \\
\textbf{OB2}&EO$_{26}$-BD$_{46}$&3.6&0.28&9.6$\pm$1 \\
\textbf{OB9}&EO$_{50}$-BD$_{55}$&5.2&0.37&10.6$\pm$1 \\
\textbf{OB18}&EO$_{80}$-BD$_{125}$&10.4&0.29&14.8$\pm$1 \\
\textbf{OB19}&EO$_{150}$-BD$_{250}$&20.0&0.28&21.0$\pm$1
\end{tabular}
\label{table1}
\end{table}

\vskip 0.5cm

\noindent
\textbf{\textsf{Materials and Methods}} \\
\textbf{\textsf{Materials}} \nolinebreak
As listed in Table \ref{table1}, a novel molecular weight series of PEO-PBD
as well as PEO-poly(ethylethylene) was synthesized by standard living anionic
polymermization techniques
\cite{hillmyer}.
The number of monomer units in each block was determined by $^1$H-NMR.
Gel-permeation chromatography with poly\-sty\-rene
standards was used to determine number-average molecular weights $\bar{M}_n$
as well as polydispersity indices (always $<$ 1.10). The
PEO volume fraction is denoted by $f_{EO}$.  For comparison, the phospholipids
SOPC (1-stearoyl-2-oleoyl phosphatidylcholine) and DMPC 
(1,2-dimyristoyl phosphatidylcholine) have hydrophilic volume
fractions $f \simeq$ 0.30 and 0.36, respectively.  \\


\noindent
\textbf{\textsf{Preparation of Polymer Vesicles}}
Giant vesicles were made by film rehydration.
Briefly, $10-50 \mu$L of a 4 mg/mL copolymer in chloroform solution was
uniformly coated on the inside wall of a glass vial, followed by
evaporation of the chloroform under vacuum for 3 h. Addition of sucrose
solution ($250-300$ mM) led to spontaneous budding of vesicles off of the
glass and into solution.
Copolymers of higher molecular weight (\textit{i.e.}, \textbf{OB18},
\textbf{OB19}) required incubation at $\simeq 60^{\circ}$C to
increase vesicle size and yield. Vesicles were usually suspended in 
phosphate-buffered saline (PBS).\\

\noindent
\textbf{\textsf{Cryogenic Transmission Electron Microscopy (cryo-TEM)}}
Thin films (about 10-300 nm) of 1.0 wt\% polymer in 
water were suspended in a microperforated grid. Samples were 
prepared in an isolated chamber with temperature and humidity 
control. The sample assembly was rapidly vitrified with liquid ethane 
at its melting temperature ($\simeq$~90 K), and kept under liquid 
nitrogen until it was loaded onto a cryogenic sample holder (Gatan 
626). Images were obtained with a JEOL 1210 at 120 kV using a nominal 
underfocus of 6 $\mu$m for improved phase contrast and digital recording. 
For a more detailed 
description and related examples, see \cite{cryo1,cryo2}. \\

\noindent
\textbf{\textsf{Optical Microscopy and Micromanipulation}}
A Nikon TE-300 inverted microscope with Narishige manipulators was used for
micropipette manipulation of vesicles.
A custom manometer system with pressure transducers (Validyne, Northridge,
CA) allowed for control and monitoring of the pressure.
Imaging was done with either a 40$\times$, 0.75 NA air objective lens
under bright-field illumination or a 20$\times$, 0.5 NA phase objective lens
for phase contrast imaging.
Bright-field imaging was used for clear visualization of the vesicle membrane,
whereas phase contrast was used when a difference in refractive indices
was established between the interior and exterior solutions (\textit{e.g.},
sucrose inside and PBS outside).
The contrast is visibly moderated by any exchange of solutes across the
membrane. 

Mechanical properties of polymersome membranes were measured by 
micropipette aspiration methods. As described previously 
\cite{discherfamily,skalak,rawicz}, a giant vesicle aspirated into a 
micropipette (of internal radius $R_p$) under an applied pressure 
$\Delta P$ leads to a projection length of membrane $\Delta L$. These 
two measured quantities are used to calculate the imposed membrane 
tension $\tau$ and the relative area dilation $\alpha \equiv \Delta 
A / A_o$ from the Law of Laplace and the outer vesicle radius $R_v$:
\begin{eqnarray}
\tau &=& \frac{1}{2} \frac{\Delta P R_v R_p}{(R_v - R_p)} \nonumber \\
\Delta A &\simeq& 2\pi R_p \Delta L \left(1 - \frac{R_p}{R_v}\right) \nonumber
\end{eqnarray}
The quantities $\tau$ and $\alpha$ are the 2-dimensional analogues
of bulk stress and strain.  
Measurements on $10-20$ individual vesicles are used to determine the
average ($\pm$ S.D.) properties of membranes reported below.\\

\begin{figure}
\epsfxsize=3.5in
\centerline{\epsfbox{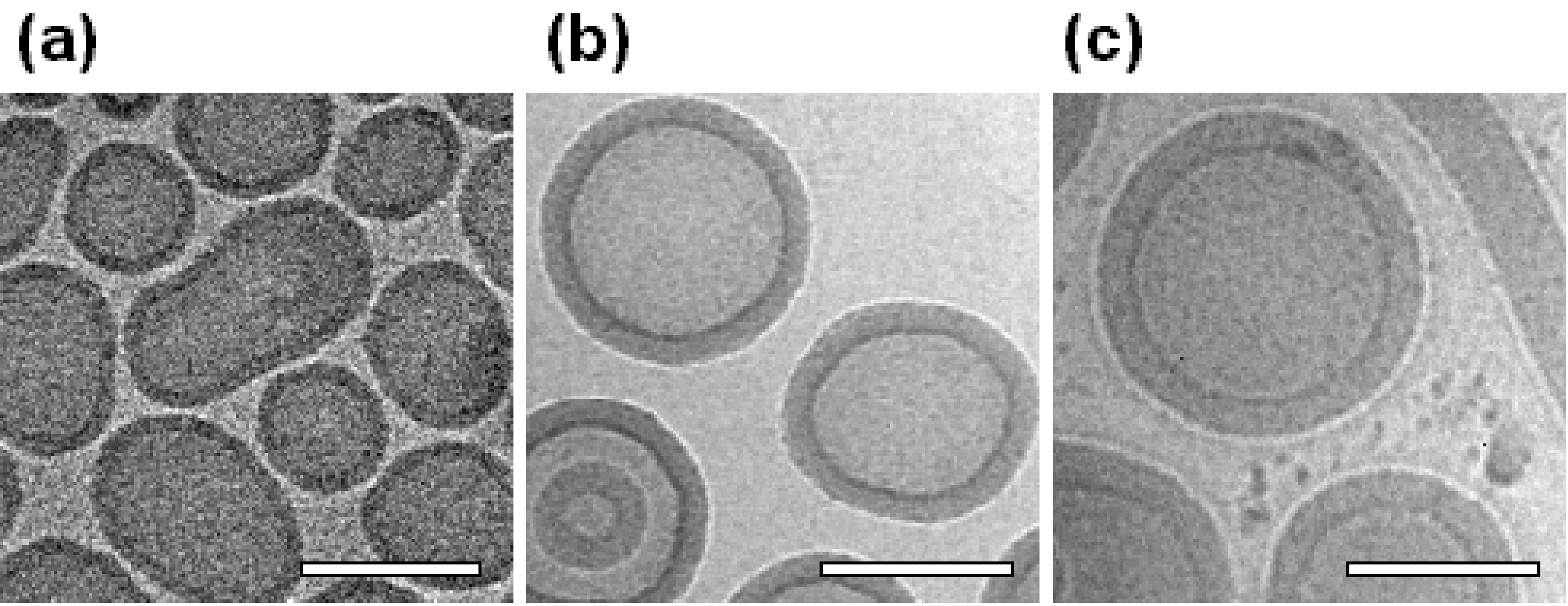}}
\caption{Cyro-TEM images of a 1.0 wt\% aqueous solution of copolymer in
water: (a) \textbf{OB2}, (b) \textbf{OB18}, and (c) \textbf{OB19}.
The hydrophobic cores of polybutadiene are the darker areas.
Scale bars are 100 nm. Polymorphism is common in cryo-TEM preparations 
\protect\cite{discherfamily} but does not 
pose any difficulties to analysis since vesicles can be clearly 
identified from their concentric-ring structure.}
\label{cryo}
\end{figure}

\noindent
\textbf{\textsf{Results and Discussion}} \\
Among all of the various vesicle-forming amphiphiles, including lipids,
a key unifying feature is a hydrophilic fraction 
$f \simeq 0.3-0.4$ (Table \ref{table1}). Aqueous self-assembly of 
the present diblocks into membranes requires such proportions, as 
it is well-documented theoretically \cite{bates} and experimentally
\cite{wonbrush} that a larger $f$ leads to wormlike and spherical micelles
while smaller values of $f$ yield inverted phases. 
Another shared feature of lipid membranes is their narrow range of 
hydrophobic core thickness $d$ of $3-5$ nm. 
Connections between molecular 
conformations and mass, as well as the interplay between these
factors in determining and limiting membrane self-assembly, thus have 
not been thoroughly studied.
Our novel series of PEO-PBD diblock copolymers allows us to 
extend the narrow range of $d$ for vesicles.  
Direct imaging of vesicles by cryo-TEM demonstrates  
a systematic increase in $d$ with $\bar{M}_n$ (Table \ref{table1} and 
Fig.~\ref{cryo}).\\

\begin{figure}
\epsfxsize=3.5in
\centerline{\epsfbox{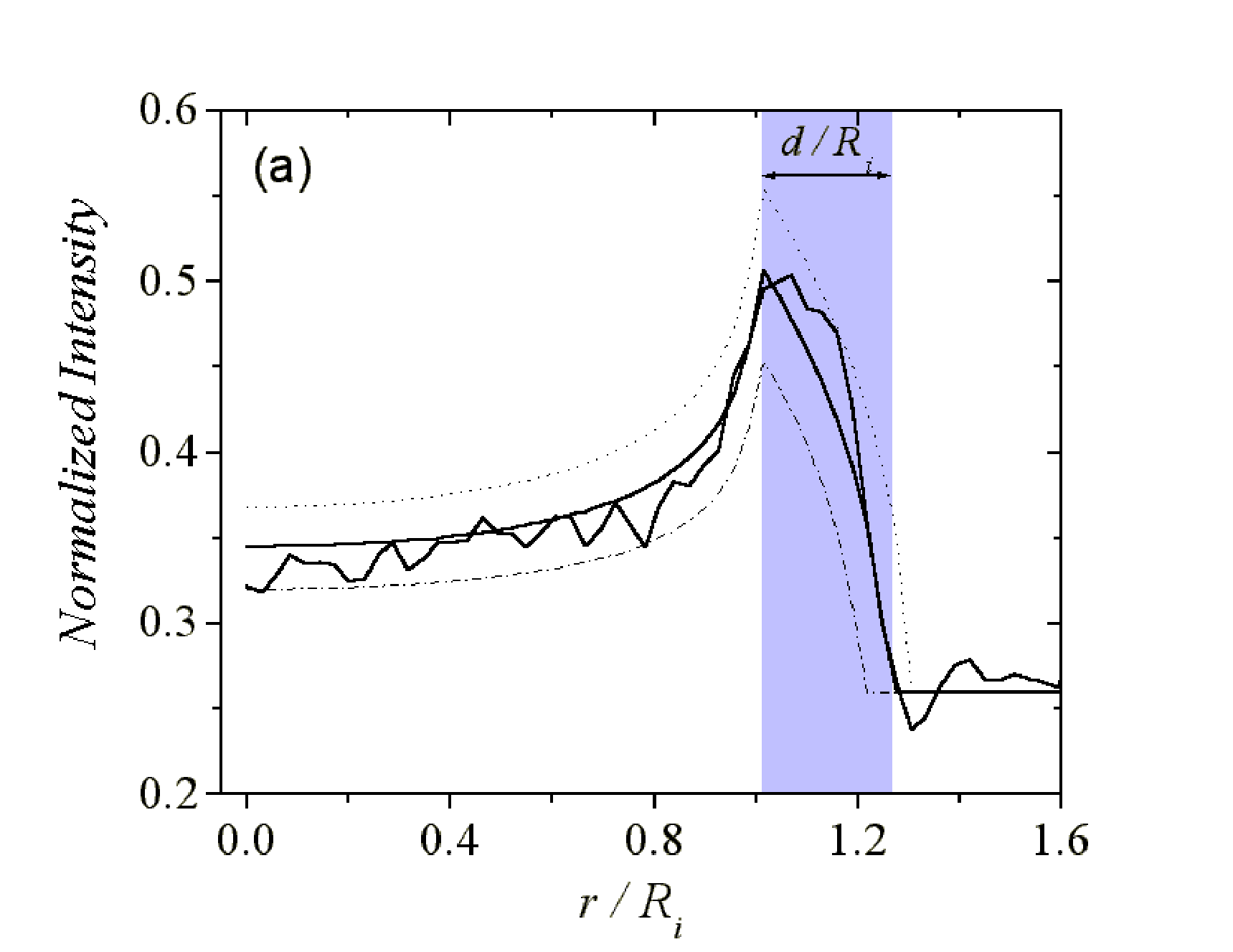}}
\epsfxsize=3.5in
\centerline{\epsfbox{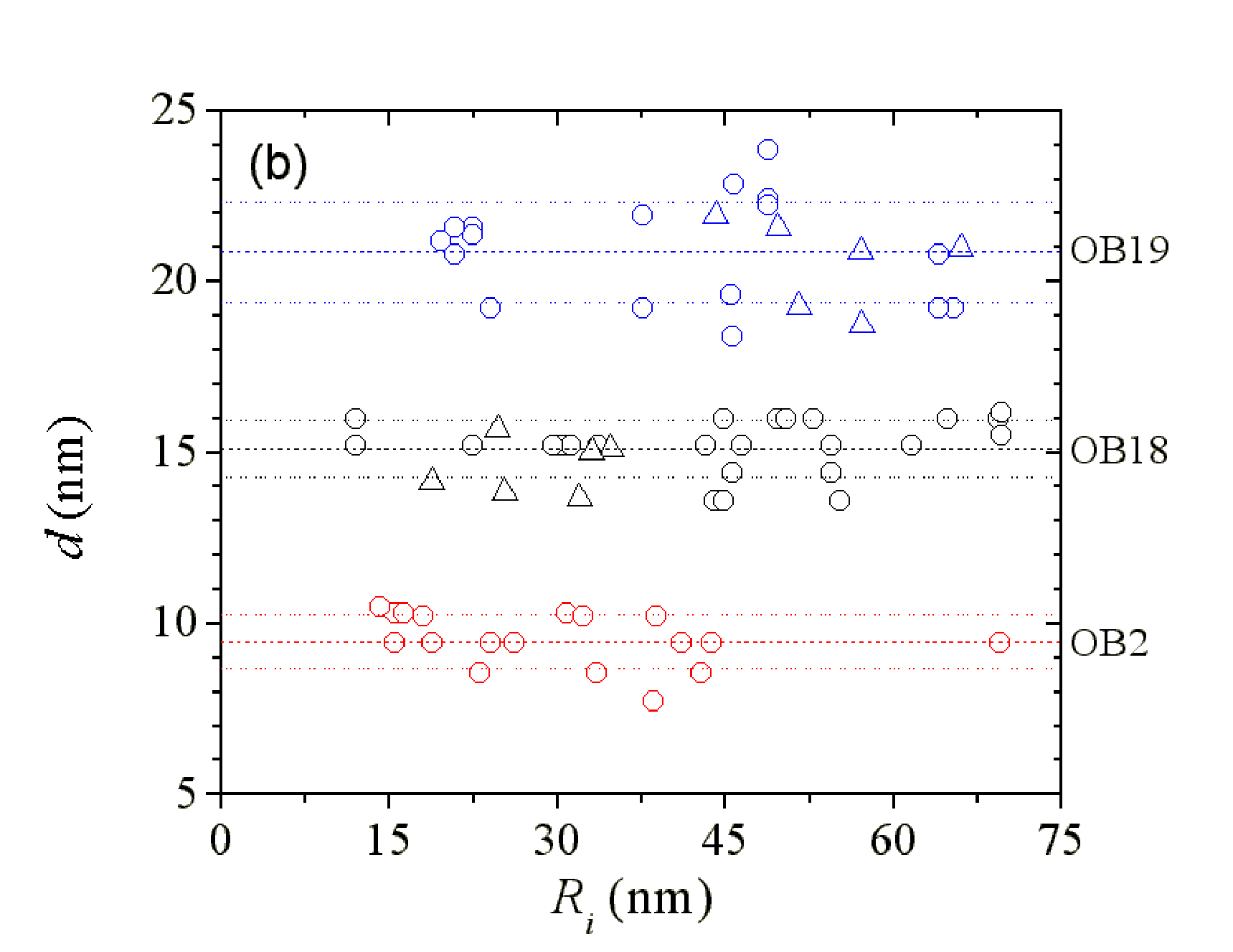}}
\caption{(a) Experimental intensity profile of a vesicle imaged by cryo-TEM.
Best fit of the data corresponds to $d/R_i=0.25$, \textit{cf.} experimental
measurement of $d/R_i=0.27$; a difference of less than 1 nm.
Dash-dot and dashed lines indicate fits using $d/R_i=0.2$ and 0.3,
respectively.  Note the Fresnel interference fringe at $r/R_i \approx 1.3$.
(b) Dependence of measured hydrophobic thickness $d$ on inner radius
$R_i$ of vesicles imaged by cryo-TEM for \textbf{OB2}, \textbf{OB18} and
\textbf{OB19}.  Solid and dashed lines are mean values $\pm$ S.D.
Data are shown using spherical vesicles ($\circ$) and 
out-of-plane curvature estimates ($\triangle$) for nonspherical vesicles.} 
\label{profile}
\end{figure}

It is important to note that
the depth-of-field for cryo-TEM is comparable to the sample film thickness
used.  As such, the resulting image is effectively the projection 
of a sample's density into a plane.  Assuming a membrane core
of homogeneous density and spherical vesicles, the projected density 
leads to a maximum in the intensity $I$ at the vesicle inner radius $r=R_i$.
At $r=R_i+d$, the intensity will go to zero, or in our case, to
the background intensity $I_o$.  This simple model for the intensity is
shown in Figure~\ref{profile}a, where $d/R_i$ is used as a free parameter.
For $d/R_i < 0.25$, the model is in excellent agreement with the
measured profile (circumferentially-averaged).  The dark and light rings
seen in Figure~\ref{cryo} are Fresnel interference fringes corresponding
to the abrupt changes in projected density at the inner and outer edges
of the membrane, respectively.  The fringes can also be seen in 
Figure~\ref{profile}a at $r \simeq R_i$ and $r \simeq R_i+d$, thus 
providing a simple means for determining the membrane thickness $d$.
Similar analysis of spherical micelles via cryo-TEM gives very comparable
results to corresponding SANS measurements \cite{wonmicelles,wonbrush}.

\vskip 0.5cm

Measurements of $d$ are shown in Figure~\ref{profile}b, based on either 
fitting experimental profiles, or otherwise, edge detection.  The results 
seem to be independent of vesicle radius even though contrast is reduced 
for smaller
vesicles. For all vesicle systems, $d$ has a standard deviation of 
$\pm 1-1.5$ nm.  

\begin{figure}
\epsfxsize=3.5in \centerline{\epsfbox{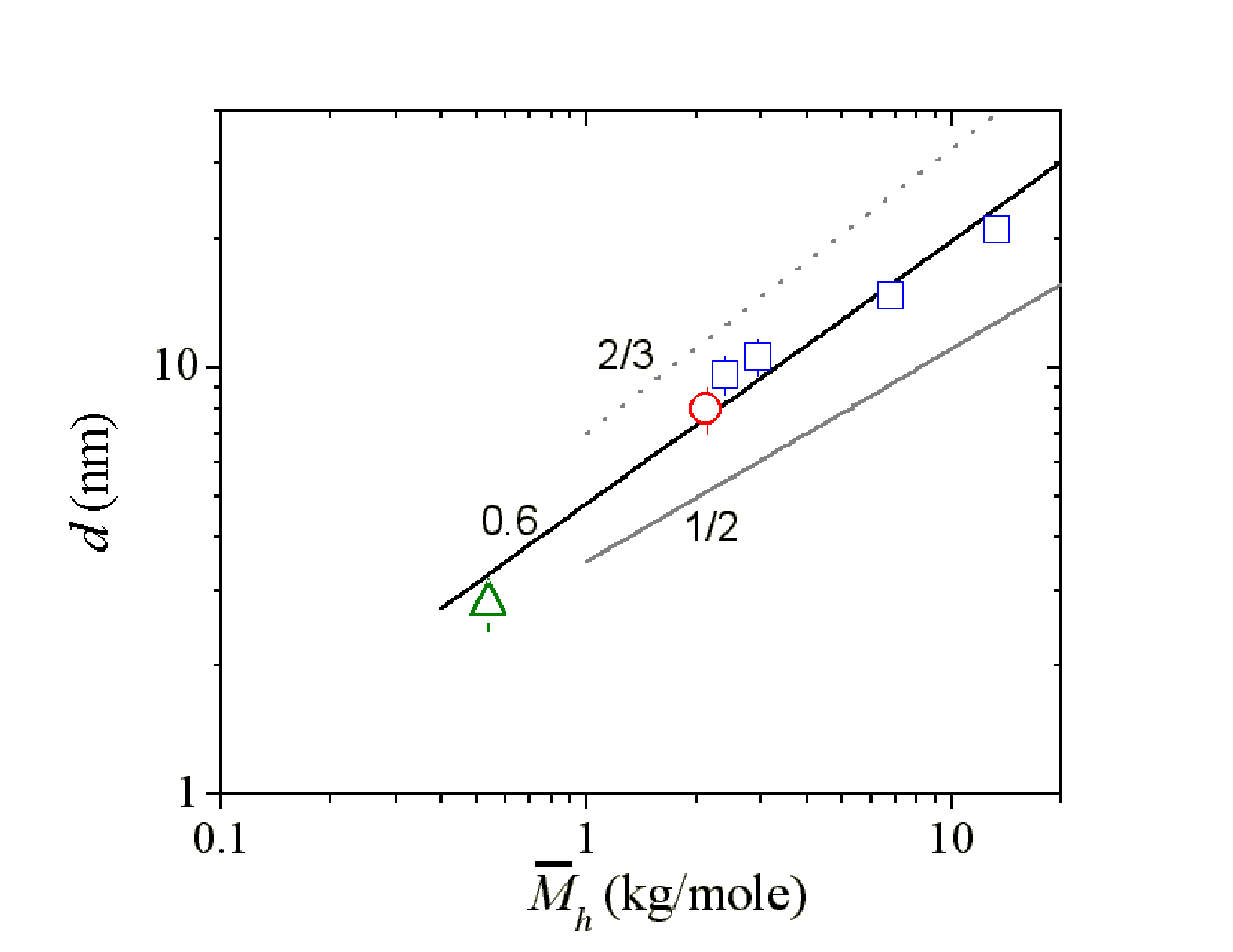}}
\caption{Scaling of core thickness $d$ with hydrophobic molecular
weight $\bar{M}_h$.
The best-fit scaling exponent of 0.6 is suggestive
of chains stretching relative to their unperturbed state but exponents of
$1/2$ and $2/3$ (solid grey and dashed grey, respectively) also fit the data
reasonably well.
Data are shown for membranes of various phospholipids ($\triangle$)
\protect\cite{marsh},
\textbf{OE7} ($\circ$), and the \textbf{OB} series ($\Box$).}
\label{d_vs_mw}
\end{figure}

Glassy diblock copolymers of PEO-poly\-sty\-rene and 
poly\-(a\-cryl\-ic acid)-poly\-sty\-rene have previously been shown to 
generate vesicular shells in organic mixtures with added water
\cite{eisenberg} but no clear relationship between copolymer 
molecular weight and membrane thickness has yet been described. The 
thickness measurements here for our self-assembling copolymers 
suggest a scaling relationship between $d$ and $\bar{M}_n$. Noting 
that the mean hydrophobic molecular weight is given by 
$\bar{M}_h \simeq \bar{M}_n (1-f)$, the experimental 
scaling of $d \sim ( \bar{M}_h ) ^a$ leads to an exponent $a \simeq$ 
0.60 (Fig.~\ref{d_vs_mw}). 
The exponent $a$ is minimally affected by including lipid data, yet
considerably expands the range of $\bar{M}_h$.
Regardless, this scaling result can offer insight into the 
chain conformations within the membrane core. In theory, fully stretched 
chains would give $a = 1$ and random coils would give $a = 1/2$.  
Our copolymers are expected to be in the strong segregation limit (SSL), 
where a balance of interfacial tension and chain entropy yield
a scaling of $a = 2/3$ \cite{bates}.
The best-fit scaling exponent therefore suggests that chains in the 
various polymersome membrane cores are stretched to some extent. 

Compared to non-equilibrated membranes of PEO-poly\-sty\-rene, \textbf{OE7} 
and \textbf{OB18} membranes have been clearly shown to be 
fluid via in-plane mobility measurements, although lateral diffusivity 
decreases strongly with $\bar{M}_n$ \cite{diffusion}. 
Fluidity generally allows for an equilibration of net forces 
that underlie the scaling exponent $a$ above, but  
there is also evidence for partial collapse of the PEO chains towards 
the interface, thus shielding the hydrophobic  
core from water \cite{wonbrush}. 
This collapse would have the effect of 
increasing the equilibrium area per chain ${\mathcal{A}}$ and decreasing 
membrane thickness, consistent with a slightly smaller exponent $a$ compared 
to the SSL.  
Assuming incompressibility, one can show that 
${\mathcal{A}} \sim (\bar{M}_h)^{0.4}$.
Additional effects associated with relatively low $\bar{M}_n$ polymers may 
also play a role in the scaling behavior. 

The high edge-contrast seen from the cryo-TEM images (Fig.~\ref{cryo})
further suggests that the interface between the hydrophobic core and the 
PEO corona is fairly narrow for all of the polymersomes.
This is consistent with theoretical predictions for copolymers in the SSL 
since interfacial thickness should scale with the Flory 
interaction parameter $\chi$ as $\sim \chi^{-0.5}$ but not with $\bar{M}_n$ 
\cite{helfand}. 
This qualitative observation on interfacial thickness 
is a first clue that the interfacial tension $\gamma$ driving the 
self-assembly of the diblocks in water is essentially constant for
this series. 

\begin{figure}
\epsfxsize=3.5in
\centerline{\epsfbox{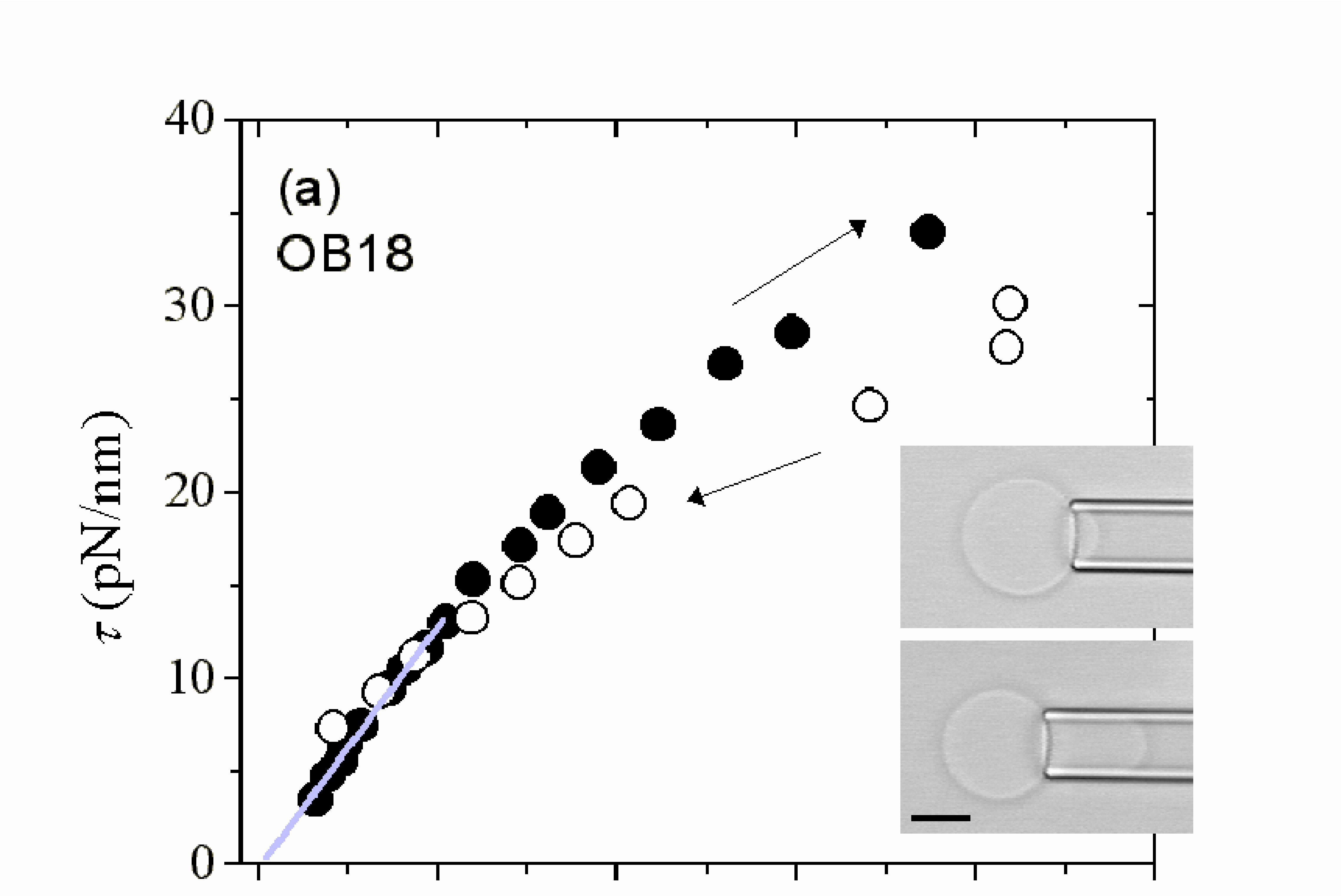}}
\epsfxsize=3.5in
\centerline{\epsfbox{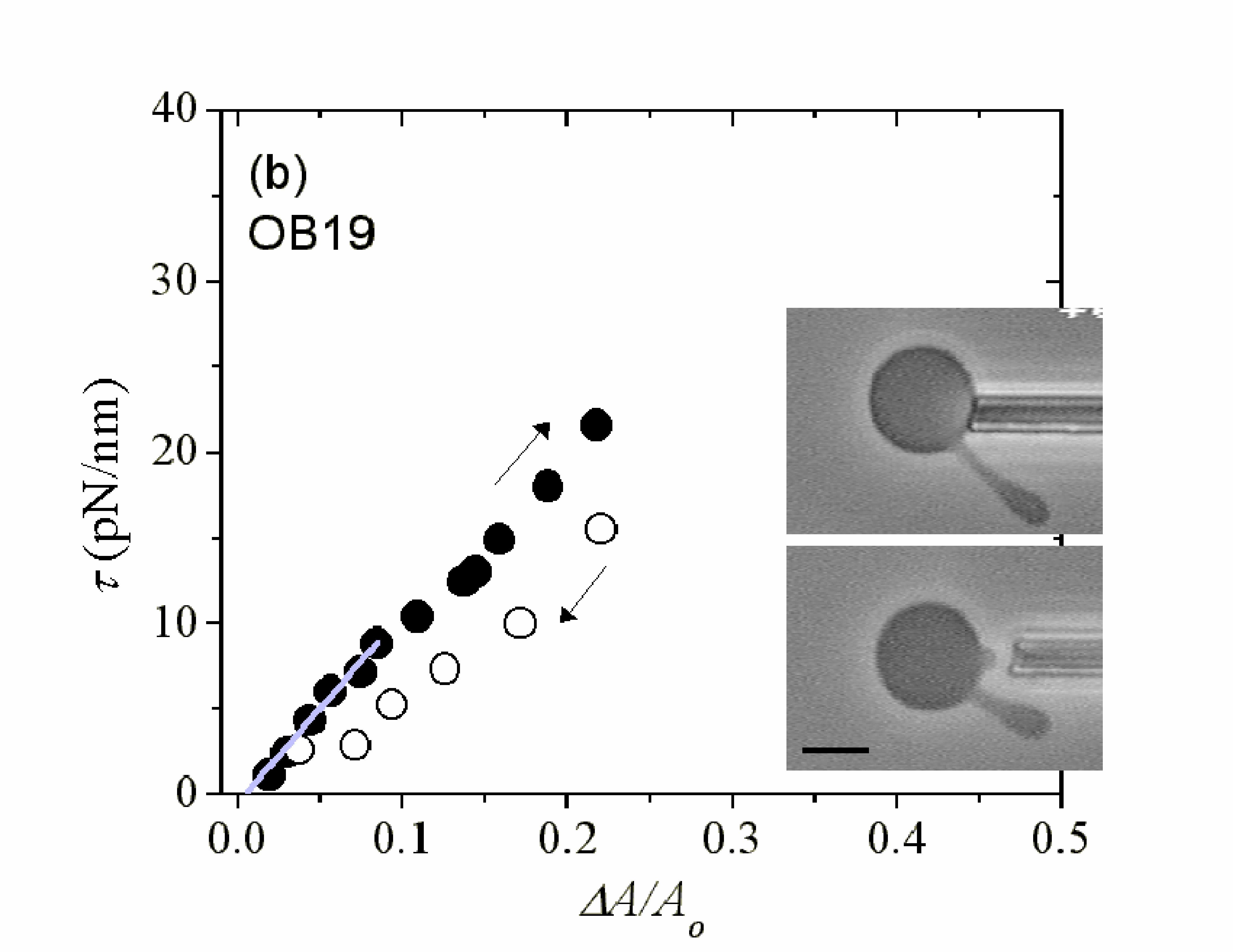}}
\caption{Determination of mechanical properties by micropipette aspiration.
The applied tension $\tau$ is plotted against areal strain $\alpha$ and the
area elastic modulus $K_a$ is determined from the initial slope.  Unlike
\textbf{OE7} vesicles whose aspiration is entirely reversible
\protect\cite{discherfamily}, hysteresis is observed after large strains
imposed upon \textbf{OB18} and \textbf{OB19} vesicles.
(a) \textbf{OB18} vesicle aspiration (\textbullet) and subsequent release
($\circ$).
The inset shows shows a typical aspiration process in bright-field imaging.
(b) \textbf{OB19} membranes do not relax
quickly to their original state. The inset shows the slow dynamics of a
twice-aspirated \textbf{OB19} vesicle under phase contrast.
In all experiments the loading rate ranged from
$\simeq 1-10 \times 10^{-4}$Nm$^{-1}$sec$^{-1}$, without significant effect
on the reported properties.  Scale bars are 5 $\mu$m.}
\label{hysteresis}
\end{figure}

Determinations of membrane elasticity and strength lend deeper insight 
into the interfacial and bulk forces at work within polymersome 
membranes. These forces have been probed by micropipette aspiration 
techniques (Fig.~\ref{hysteresis}) pioneered by Evans and 
coworkers \cite{skalak} with giant unilamellar lipid vesicles. 
Plots of the 
effective membrane tension $\tau$ against the mean dilational strain 
$\alpha$ reveal an initial linear response as well as
subsequent nonlinear and hysteretic effects. The latter are 
obvious for the thicker membranes \textbf{OB18} and \textbf{OB19} 
at areal strains ($>$10-20\%) much greater than those 
sustainable by any lipid membrane. 
Nonlinear behavior and hysteresis are thus not accessible with any 
of the various elastic lipid systems. 
Nevertheless, the reproducible initial slope of $\tau$ versus $\alpha$ 
defines an area elastic modulus $K_a$ for the membrane 
(Fig.~\ref{hysteresis}). 

Only one series of single component phospholipid membranes (consisting of 
saturated and unsaturated phosphatidylcholines) has been thoroughly 
characterized.  The most recent and refined measurements give 
$K_a \simeq$ 240 pN/nm \cite{rawicz}.
The considerable thermal undulations of lipid membranes complicate 
$K_a$ measurements, requiring a significant correction to 
account for the entropic contributions to area dilation.  For our
polymersomes this effect is mitigated by membranes that are
the substantially thicker, and hence stiffer out-of-plane; for \textbf{OE7}, 
the correction to $\alpha$ is only about 1\%.

In addition to hydrophobic interactions, 
other factors affecting $K_a$ can arise from the 
counterion pairing expected among zwitterionic amphiphiles or 
the presence of small molecules, such as cholesterol, in the membrane.
These complexities are avoided by the use of single-component, neutral
systems such as those here.  
Hence we can essentially view $K_a$ as being primarily related to the 
interfacial tension $\gamma$ that 
reflects the chemical composition at each interface of the membrane.  A 
simple area elasticity calculation \cite{israel} based on balancing 
molecular compression ($\sim 1/{\mathcal{A}}$) against interfacial energy 
($\sim {\mathcal{A}} \gamma$) gives $K_a = 4\gamma$. 

\begin{figure}
\epsfxsize=3.5in
\centerline{\epsfbox{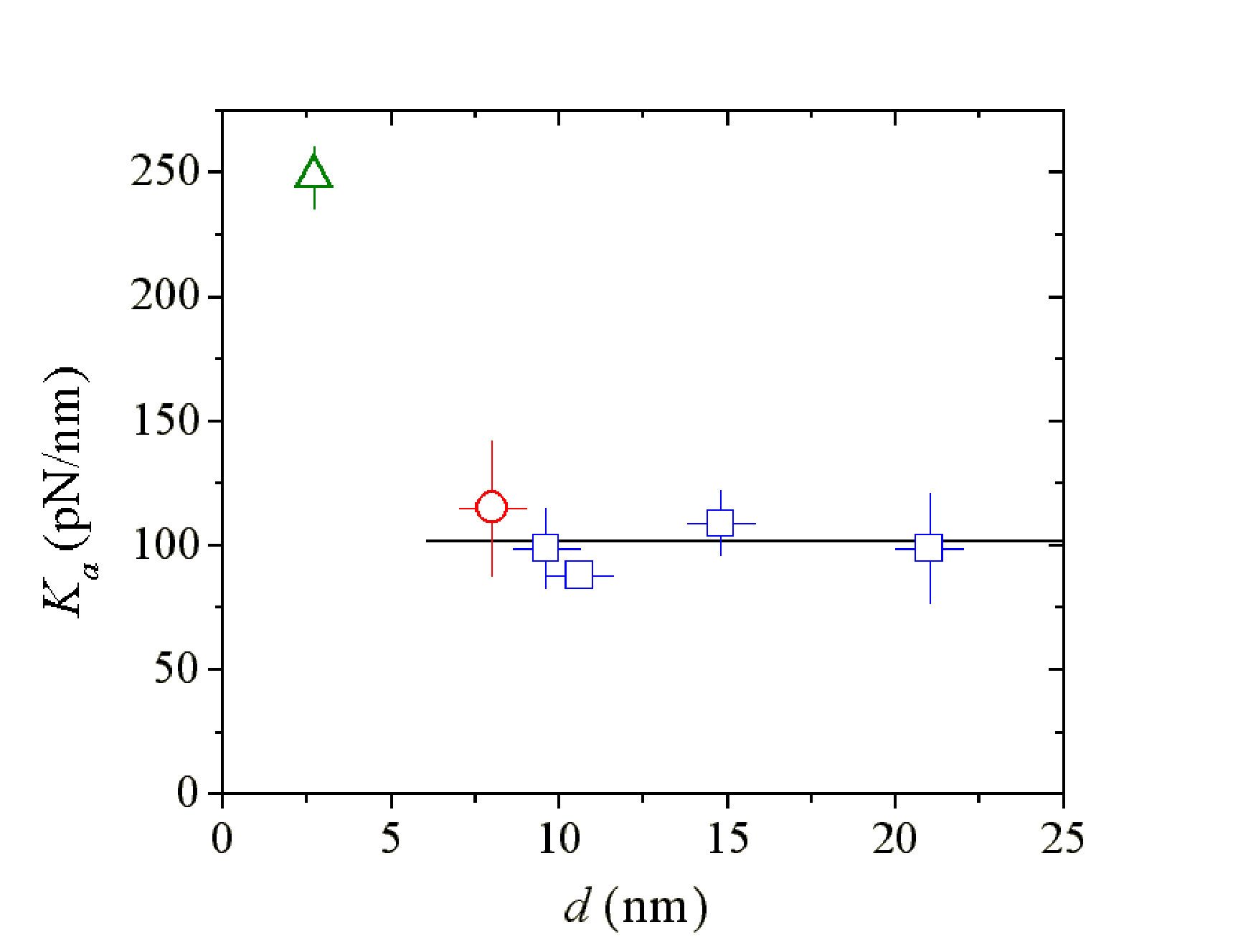}}
\caption{Molecular weight independence of the area elastic modulus $K_a$.
The dominant factor in determining $K_a$ is the
interfacial tension $\gamma$ or equivalently, the interaction
parameter $\chi$ that drives segregation.  The membrane elasticity is
thus determined strictly by the chemical composition of the interface
and not the size of the molecule.
Membranes of phosphatidylcholines (PC) have somewhat higher values of $K_a$
\protect\cite{rawicz}, reflecting the distinct composition.
Data are shown for various PC ($\triangle$) \protect\cite{rawicz},
\textbf{OE7} ($\circ$), and \textbf{OB}
($\Box$) vesicles.
Mean $K_a$ is $102 \pm 10$ pN/nm.}
\label{ka_vs_d}
\end{figure}

The chemical rather than physical basis for $\gamma$ leads 
one to expect that $K_a$ is independent of $\bar{M}_n$ (and hence 
$d$). Indeed a mean $K_a$ of 102 $\pm$ 10 pN/nm is obtained for all 
of the various polymersomes
(Fig.~\ref{ka_vs_d}).  This includes \textbf{OE7}, which is simply a 
hydrogenated \textbf{OB}.  Surface elasticity of the membrane thus 
depends
\emph{only} on the interface.  Moreover, enthalpic interactions between PEO
chains, which have been speculated to include H$_2$O bridging 
\cite{fuller} or crystallization \cite{hillmyer} are either 
independent of PEO length or simply not a factor. A value of 
$\gamma=K_a/4=26$ pN/nm is also very typical of oil-water interfaces. 
As mentioned, $\gamma$ and $\chi$ are related and provide a measure 
of segregation between blocks; specifically, $\gamma \sim \sqrt{\chi}$ 
\cite{helfand}.  The results thus suggest that a combined 
knowledge of amphiphile geometry (\textit{i.e.}, $f$) and 
interaction energies ($\chi$) lead to predictive insights into 
membrane structure and elasticity.

\begin{figure}
\epsfxsize=3.5in
\centerline{\epsfbox{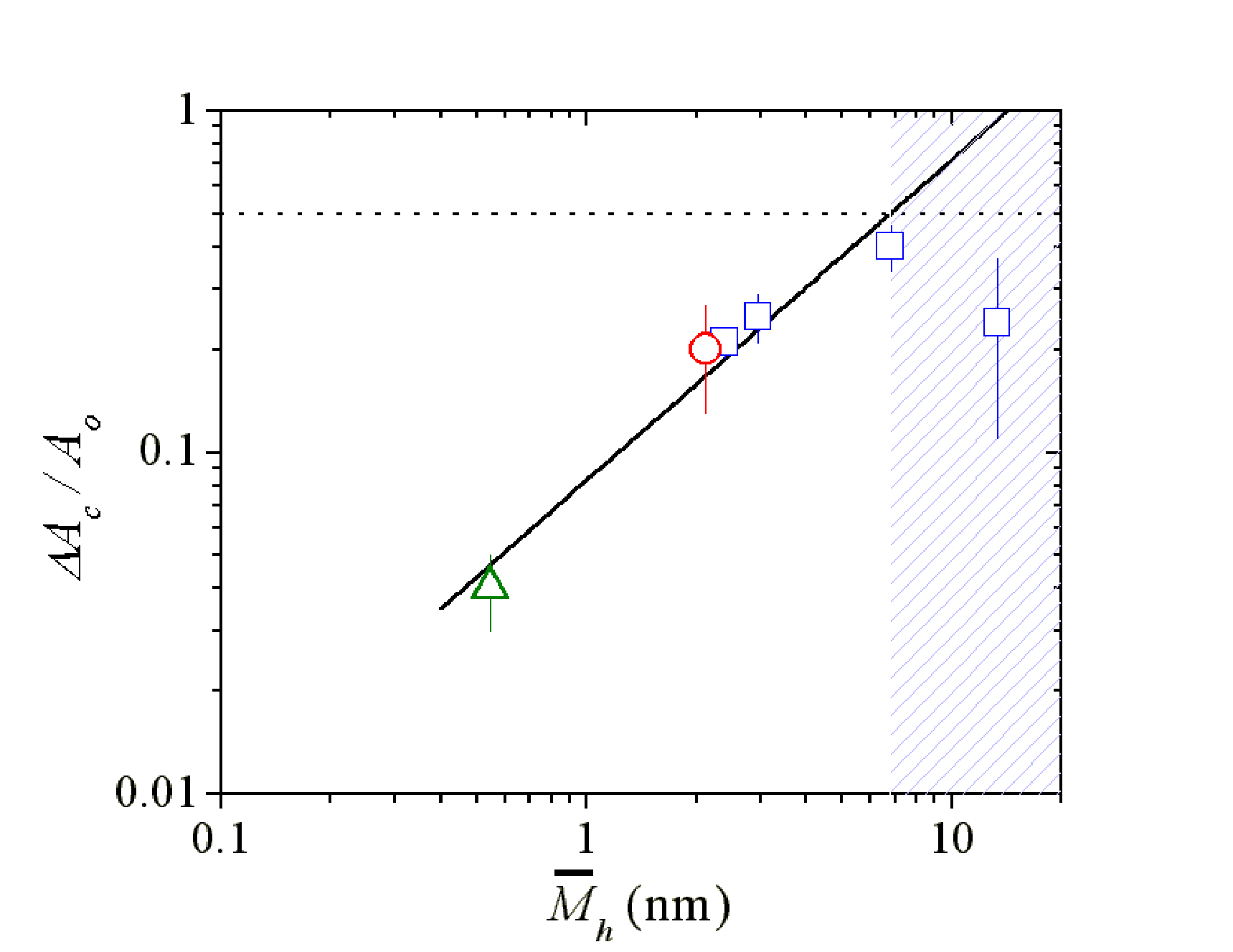}}
\caption{Areal strain at rupture $\alpha_c$ versus hydrophobic molecular 
weight.  Scaling
behavior is almost that of a simple linear dependence with an exponent
$\simeq$ 0.94.
Data are shown for SOPC ($\triangle$), \textbf{OE7} ($\circ$), and
\textbf{OB} ($\Box$) vesicles.
The predicted upper limit of 50$\%$ is universal to bilayer systems.  Note
that lipid membranes are well below this bound and typically do not exceed
$\simeq$~5$\%$ areal strains.  The hatched region schematically illustrates
the molecular weight range where chain entanglements are thought to contribute
\protect\cite{ferry}.}
\label{a_vs_d}
\end{figure}

While phosphatidylcholine membranes appear slightly stiffer than 
the present polymersome membranes, no lipid membrane can be strained 
by more than a critical strain $\alpha_c \simeq$~5$\%$ before rupture, 
regardless of cholesterol addition. In 
contrast, the present synthetic systems can be strained to almost 
50$\%$, with a nearly linear dependence on molecular weight 
(Fig.~\ref{a_vs_d}).  At such large strains, an incompressible 
membrane will thin considerably to a
reduced thickness $d_c \equiv d/(1+\alpha_c)$. 
Using the previous relation $d \sim {(\bar{M}_h)}^a$ gives
the scaling $\alpha_c \sim {d_c}^b$ with $b \simeq$ 1.7. 
This scaling excludes the largest copolymer, \textbf{OB19}, 
which generally exhibits the earliest onset of hysteresis and falls 
well below the trend (Fig.~\ref{a_vs_d}). 
As explained below, the apparent $\tau_c$ and 
$\alpha_c$ are both smaller for \textbf{OB19} 
($\tau_c=22 \pm 5~$pN/nm) than for \textbf{OB18} ($\tau_c=33 \pm 5~$pN/nm). 
Thus, although larger copolymers allow for larger areas per chain 
${\mathcal{A}}$, there are upper bounds on the strain 
(and stress) that can be withstood by a membrane.

The same balance of forces used to understand the SSL and membrane elasticity
provides insight into membrane stability limits. The net chain pressure 
$\Pi$ (core 
plus headgroup) and applied tension $\tau$ are balanced by the 
interfacial tension $\gamma$:
\noindent
\begin{eqnarray}
\Pi + \tau = 2 \gamma \label{simple}
\end{eqnarray}
To account for the nonlinearity in the aspiration plots of
Figure~\ref{hysteresis}, the isotropic membrane tension is expanded 
\cite{skalak} to second order:
\begin{equation}
\tau(\alpha) = \tau_0 + K_a \alpha + {1 \over 2} \frac{\partial ^3F}
{\partial \alpha^3} \alpha^2
\end{equation}
Because of isotonic conditions, $\tau_0=0$, and the experiments are well-fit by
\begin{equation}
\tau(\alpha) = K_a(\alpha - c \alpha^2)
\label{empirical}
\end{equation}
with the coefficient $c \equiv -K_a^{-1}({1 \over 2}
\partial^3F/\partial\alpha^3)$ having the average value of 1.0 $\pm$ 0.2 for
\textbf{OB9}, \textbf{OB18} and \textbf{OB19}.
Using the previously cited mean-field result of $K_a=4 \gamma$, we obtain
\begin{eqnarray}
\Pi + K_a (\alpha - \alpha^2) = \mbox{${1 \over 2}$}K_a
\end{eqnarray}
and solve for $\alpha$ to arrive at
\begin{eqnarray}
\alpha = \mbox{${1 \over 2}$}\left(1 \pm \sqrt{ \Pi / \gamma - 1} \right)
\label{soln}
\end{eqnarray}

\noindent
From Eq.~(\ref{soln}), there can only be real solutions provided that
$ \Pi \geqslant \gamma$.  Noting that $\Pi = 2\gamma$ at zero applied tension,
$\gamma \leqslant \Pi \leqslant 2\gamma$.  Establishing the
bounds for $\Pi$ allows us to do the same for $\tau$  via Eq.~(\ref{simple})
such that $\gamma \geqslant \tau \geqslant 0$.  The upper bound for $\tau$ 
could also have been obtained by setting $d\tau / d\alpha = 0$ 
from Eq.~(\ref{empirical}).

By definition, $\alpha \geqslant 0$, but solutions of Eq.~(\ref{soln}) with the
positive root give $\mbox{${1 \over 2}$} \leqslant \alpha \leqslant 1$ whereas
those with the negative root give 
$0 \leqslant \alpha \leqslant \mbox{${1 \over 2}$}$.
Only the latter makes physical sense, corresponding to $\Pi / \gamma = 1-2$
and $\tau/\gamma=1-0$.
The above bounds of $\alpha \leqslant \mbox{${1 \over 2}$}$ and
$\tau \leqslant \gamma$ largely agree with the experimentally
observed limits self-assembled polymersome membranes.  
A related case where the core polymer 
is treated as a three-dimensional brush \cite{rawicz} would give 
$\alpha \leqslant 0.21$, which is
exceeded here. The overall membrane behavior also appears rather 
insensitive to any local variations associated with finite 
polydispersity and seems instead dominated by the collective behavior 
of a fluid or melt-like state.  Thus the increased thickness (\textit{i.e.}, 
larger $\bar{M}_n$) makes the interface more readily self-healing. In 
natural membranes, by comparison, stiffening and toughening of
the membranes is mediated by the small molecule cholesterol - 
presumably through cohesive healing of defects. 
However, the additional stability imparted by cholesterol to biomembranes 
cannot compare with that of a much thicker membrane. 
The results here therefore imply
that biomembranes are not designed for maximal stability, but are instead
optimized for a balance between stability and fluidity.

The nonlinearity seen in the stress-strain curves, $\tau(\alpha)$, 
also appears distinctive and revealing. 
As already noted, lipid membranes cannot withstand areal strains 
exceeding $\simeq$~5$\%$ and therefore a strictly linear 
elastic response is not surprising.  
For such systems, the corresponding first-order analysis ($c=0$)
of the stabililty limit 
again yields $\alpha \leqslant \mbox{${1 \over 2}$}$ (independent of 
$\gamma$), although the additional conditions of 
$\Pi/\gamma \geqslant 1$ and $\tau/\gamma \leqslant 1$ would not be 
apparent.
The basis for the strain-softening seen here is not 
clear.  The nonlinearity is not strongly dependent on deformation rate, 
suggesting that this is not a collective process involving many molecules
but is instead a 
rearrangement at the molecular scale. We speculate that area dilation 
decreases PEO stretching and allows more collapse and hence shielding of 
the hydrophobic core. The proposed process is inspired in part by compressed 
monolayers which tend to show a decreased slope in their pressure 
isotherms during large dilations.

The decreased stability of the thickest membrane (\textbf{OB19}) is 
also unclear at this point but seems likely to be the result of increasing 
physical entanglements between chains. The inset to 
Figure~\ref{hysteresis}b is representative of the very slow relaxation 
dynamics of \textbf{OB19} membranes. Even in 
\textbf{OB18}, membrane dynamics following electroporation are more 
than 100 times slower than \textbf{OE7} dynamics \cite{helim}. 
Furthermore, lateral diffusion coefficients
beginning with \textbf{OB18} exhibit activated reptation 
\cite{diffusion}, which is a much stronger function of $\bar{M}_{n}$ 
than simple Rouse diffusion.

Provided that the timescale for aspiration is much smaller than the 
timescale for rearrangement among polymer chains (as is likely at the 
largest $\bar{M}_n$), the entanglements present could act in a similar 
way to covalent crosslinks. Surprising perhaps, but consistent with the 
results here, polymersome membranes with very low crosslink densities 
have been found to be weaker than uncrosslinked membranes \cite{xlink}.  
This destabilization presumably arises through stress localization; 
that is, the tension $\tau$ is inhomogeneous over the membrane due to 
slow relaxations that oppose equilibration of forces. 

Non-equilibrium effects indicated above can also be seen in 
$\tau$-$\alpha$ hysteresis loops following graded release from 
aspiration (Fig.~\ref{hysteresis}).  Even down to low apparent areal 
strains of less than 10\%, \textbf{OB19} exhibits marked hysteresis, 
whereas \textbf{OB18}
aspiration appears reversible up to more modest strains of $\simeq 10-15\%$. 
In contrast, aspiration of \textbf{OE7} is reversible for nearly 
all strains up to lysis \cite{discherfamily}, consistent with diffusion studies 
indicating Rouse-type mobility \cite{diffusion}. 
Hence the hysteretic behavior in thicker membranes likely reflects 
relaxation times that scale strongly \cite{lodge} with $\bar{M}_{n}$.\\

\noindent
\textbf{\textsf{Conclusions}} \\
Vesicles formed by superamphiphiles provide new insight into some of 
the basic properties of bilayer membranes. 
By use of synthetic diblock copolymers, limitations of previous 
membrane systems have been considerably exceeded, providing novel insights
into structure, scaling and physical limits on lamellae. 
Specifically, the surface elasticity is found to be scale-independent, in 
accordance with simple mean-field theories. 
The membrane lysis tension $\tau_c$ and critical areal strain $\alpha_c$ 
are found to increase with $\bar{M}_n$, but only up to a simple limit.  The 
onset of chain entanglements with higher $\bar{M}_n$ 
introduces novel bulk effects that eventually undermine 
interfacial elasticity through slowed response times.  
Examination of membranes assembled from PEO-PBD-PEO triblocks, where 
linear and looped configurations 
are expected, may help clarify such mechanisms. Of additional 
interest will be determinations of other properties such as the 
bending modulus which is expected to scale as $\sim K_a d^2$ 
\cite{discherfamily} for interface-dominated membranes. 
Finally, while it is clear that lipid membranes found in nature are not
maximally stable, they have developed sufficient stability while also 
providing the fluidity necessary for diverse functions. \\

\noindent
{\small
This work was supported by NSF-MRSEC's at Penn and University of Minnesota
as well as a materials science grant from NASA.  HB thanks Dr. H.
Aranda-Espinoza at Penn for many valuable conversations.  The authors also
acknowledge useful discussions with Prof. E.A. Evans at Boston University.}

%
%

\end{document}